\documentclass[review]{elsarticle}

\usepackage{lineno,hyperref}
\usepackage[nolist]{acronym}
\usepackage[onehalfspacing]{setspace}
\usepackage{booktabs}
\usepackage{float}
\usepackage{amsmath}
\usepackage{listings}
\modulolinenumbers[1]

\journal{Astroparticle Physics}

\begin{document}

\begin{frontmatter}

\title{Bokeh Mirror Alignment\\ for Cherenkov Telescopes}
\author[a]{M.~L.~Ahnen}
\author[d]{D.~Baack}
\author[b]{M.~Balbo}
\author[c]{M.~Bergmann}
\author[a]{A.~Biland}
\author[c]{M.~Blank}
\author[a]{T.~Bretz}
\author[d]{K.~A.~Bruegge}
\author[d]{J.~Buss}
\author[d]{M.~Domke}
\author[c]{D.~Dorner}
\author[d]{S.~Einecke}
\author[c]{C.~Hempfling}
\author[a]{D.~Hildebrand}
\author[a]{G.~Hughes}
\author[a]{W.~Lustermann}
\author[c]{K.~Mannheim}
\author[a]{S.~A.~Mueller\corref{cor1}}
\author[a]{D.~Neise}
\author[b]{A.~Neronov}
\author[d]{M.~Noethe}
\author[d]{A.-K.~Overkemping}
\author[c]{A.~Paravac}
\author[a]{F.~Pauss}
\author[d]{W.~Rhode}
\author[a]{A.~Shukla}
\author[d]{F.~Temme}
\author[d]{J.~Thaele}
\author[b]{S.~Toscano}
\author[a]{P.~Vogler}
\author[b]{R.~Walter}
\author[c]{A.~Wilbert}
\cortext[mycorrespondingauthor]{S. A. Mueller\\sebmuell@phys.ethz.ch}
\address[a]{ETH Zurich, Institute for Particle Physics\\
Otto-Stern-Weg 5, 8093 Zurich, Switzerland}
\address[b]{University of Geneva,  ISDC Data Center for Astrophysics\\
 Chemin d'Ecogia 16,  1290 Versoix,  Switzerland}
\address[c]{Universit\"at W\"urzburg, Institute for Theoretical Physics and Astrophysics\\
Emil-Fischer-Str. 31, 97074 W\"urzburg,  Germany}
\address[d]{TU Dortmund, Experimental Physics 5\\
Otto-Hahn-Str. 4, 44221 Dortmund, Germany}
\begin{abstract}
\acfp{iact} need imaging optics with large apertures and high image intensities
to map the faint Cherenkov light emitted from cosmic ray air showers onto their
image sensors.
Segmented reflectors fulfill these needs, and composed from mass production mirror facets they are inexpensive and lightweight.
However, as the overall image is a superposition of the individual facet images, alignment remains a challenge.
Here we present a simple, yet extendable method, to align a segmented reflector using its Bokeh.
Bokeh alignment does not need a star or good weather nights but can be done even during daytime.
Bokeh alignment optimizes the facet orientations by comparing the segmented reflectors Bokeh to a predefined template.
The optimal Bokeh template is highly constricted by the reflector's aperture and is easy accessible.  
The Bokeh is observed using the out of focus image of a near by point like light source in a distance of about $10$ focal lengths.
We introduce Bokeh alignment on segmented reflectors and demonstrate it on the  \ac{fact} on La Palma, Spain.
\end{abstract}
\end{frontmatter}
\newcommand{\FigCapLabSca}[4]{
\begin{figure}[H]
    \begin{center}
        \includegraphics[width=#4\textwidth]{#1}
        \caption[]{#2}
        \label{#3}
    \end{center}
\end{figure}
}
\newcommand{\FigCapLab}[3]{
    \FigCapLabSca{#1}{#2}{#3}{1.0}
}
%
\newcommand{\OutOfFocus}{\textit{out of focus}}
\newcommand{\InFocus}{\textit{in focus}}
\newcommand{\FocalRatio}{F}
\newcommand{\FocalLength}{f}
\newcommand{\ImageDistance}{b}
\newcommand{\ImageSensorDistance}{d}
\newcommand{\ObjectDistance}{g}
\newcommand{\ApertureDiameter}{D}
\newcommand{\ApertureFunction}{A}
\newcommand{\BokehFunction}{B}
\newcommand{\BokehRadius}{r_\BokehFunction}
\newcommand{\BokehTemplateFunction}{\BokehFunction_{\text{template}}}
\newcommand{\ApertureRadius}{r_\ApertureFunction}
%
\section{Introduction}
The \ac{iact} technique and its large effective area has opened the very high energy $\gamma$ ray sky to astronomy.
Almost \cite{CANGAROO1_optics} all former \cite{WHIPPLE_optics, CAT_Themis_optics}, current \cite{VERITAS_optics, HESS_I_optics, HESS_II_optics, MAGIC_optics, FACT_design}, and future \acsp{iact} \cite{CTA_Introduction, TAIGA_IACT_optics} make use of segmented imaging reflectors with apertures up to $614\,$m$^2$ \cite{, HESS_II_optics}.
Segmented reflector facets can be mass produced inexpensively with an acceptable image quality.
Facets are much lighter than a monolithic mirror, and thus allow for very fast telescope repositioning e.g. for $\gamma$ ray burst hunting.\\
However, there is one challenge to segmented reflectors.
This is the task of manipulating the mirror facet orientations and positions to improve the image quality, also known as alignment.
Alignment needs to be done not only during installation but also in case of repair and replacement of facets.
To find the few $\gamma$ ray induced events in the far more numerous class of hadronic cosmic ray induced events, the air shower records are analyzed for geometrical properties.
This makes imaging quality and alignment important for an \ac{iact}.\\
\section{Current methods}
\label{secCurrentAlignmentMethods}
To tackle the challenge of alignment, several approaches are in use.
We can summarize them in three categories.
%
First, there is the $2\FocalLength$ alignment, which is simple and does neither need star light nor dark nights \cite{CAT_Themis_optics, FACT_design,VERITAS_bias_alignment}.
However, $2\FocalLength$ alignment is geometrically very restricted, which makes e.g. reaching the inner facets, that might be shadowed by the image sensor housing, a challenge.
A second class of alignment methods uses remote controllable orientation actuators on the mirror facets and the live image of a star on a dedicated screen \cite{HESS_II_optics, MAGIC_amc}.
The screen is best positioned in the reflector's focal plane, which might conflict with the image sensor position.
Since the mirror actuation needs to be used extensively to overcome the \acs{psf} facet ambiguity, these methods do not scale to well with the number of facets.
Also a lot of custom hardware is needed.
Furthermore one gains no additional information about the individual facet surfaces.
Most importantly this method needs a bright star and can only tolerate faint clouds. 
Only if the image sensor distance can be altered or the out of focus effect can be corrected for differently, then a close by artificial light source can be used instead of a star.
Third, there are methods based on the ideas of \acs{sccan} \cite{SCCAN_Arqueros_2005}, which track a star and observes the reflections of the facets on a camera located in the reflector's focal point \cite{VERITAS_SCCAN_alignment, FACT_NAMOD_alignment}.
However, this method is also complex, and needs a bright star.
Without a second camera, the method also favors clear nights \cite{FACT_NAMOD_alignment}.
Although the mounting of the camera in the reflector's focal point can be overcome with a $45^\circ$ mirror, no \acs{sccan} based \acs{iact} alignment implementation \cite{VERITAS_SCCAN_alignment, FACT_NAMOD_alignment} is mounted permanently yet.
\newline
In this paper we present a new method, that overcomes most of these limitations.
We call it Bokeh alignment.
This novel method is less geometrically restricted than the $2\FocalLength$ method and has fewer problems regarding shadowing.
It is a simple method, which does not need a lot of custom hardware or software.
Yet it gives information about individual mirror facet surfaces. 
Bokeh alignment does not need equipment to be installed in the reflector's focal point or focal plane.
Finally it can be done anytime, even during the day.
\newline
We will first introduce the Bokeh and the Depth of Field on general, thin imaging systems.
Second, we will discuss the Bokeh in the special case of segmented reflectors and its close relation to alignment.
Third, we will present how to align a segmented reflector using Bokeh alignment without star light or dark nights.
Finally, we will show the most simple flavor of Bokeh alignment, done on the \acf{fact} in a step by step manner.
%
\section{Depth of Field and the Bokeh}
All real imaging optics have a non zero aperture, i.e. the aperture diameter $\ApertureDiameter$ is greater zero.
Furthermore, all non zero aperture imaging optics have the effect of Depth of Field. 
The Depth of Field makes the image of objects look blurred on the image sensor in case these objects are \OutOfFocus{}.
This blurring is described by the Bokeh function $\BokehFunction$ of the imaging system.
Bokeh is best known in photography where it is used to achieve certain image compositions \cite{Bokeh_Photography_Merklinger_1997}.
The Depth of Field blurring has also been discussed for \acs{iact}s \cite{IACT_depth_of_field, How_to_focus_Cherenkov_telescopes_Hoffmann, CTA_Monte_Carlo_study_2013}, in particular in the context of image and object distances and its impact on the air shower image analysis.
%
\subsection{Bokeh on thin imaging optics}
\label{SecBokehThin}
The idealized thin imaging optic bends the incoming light only in its aperture plane. 
In practice, this is a good approximation for lenses and reflectors with a large $\FocalRatio$-number, $\FocalRatio = \FocalLength/\ApertureDiameter$ where $\FocalLength$ is the optic's focal length.
On thin imaging optics, the Bokeh effect is described by the geometry of the thin lens equation
\begin{eqnarray}
    \label{EqThinLens}
    \frac{1}{\FocalLength} &=& \frac{1}{\ImageDistance} + \frac{1}{\ObjectDistance},
\end{eqnarray}
together with an image sensor plane located at distance $\ImageSensorDistance$, see Figure \ref{FigThinLensBokeh}.
The thin lens equation defines the image distance $\ImageDistance$ where an image sensor must be placed, on an imaging system with focal length $\FocalLength$, in order to obtain a sharp image of an object in distance $\ObjectDistance$.
In contrast to this \InFocus{} situation, the image sensor can be placed elsewhere, e.g. in $\ImageSensorDistance \neq \ImageDistance$, and the image of the object in distance $\ObjectDistance$ becomes \OutOfFocus, see Figure \ref{FigThinLensBokeh}.
\FigCapLabSca{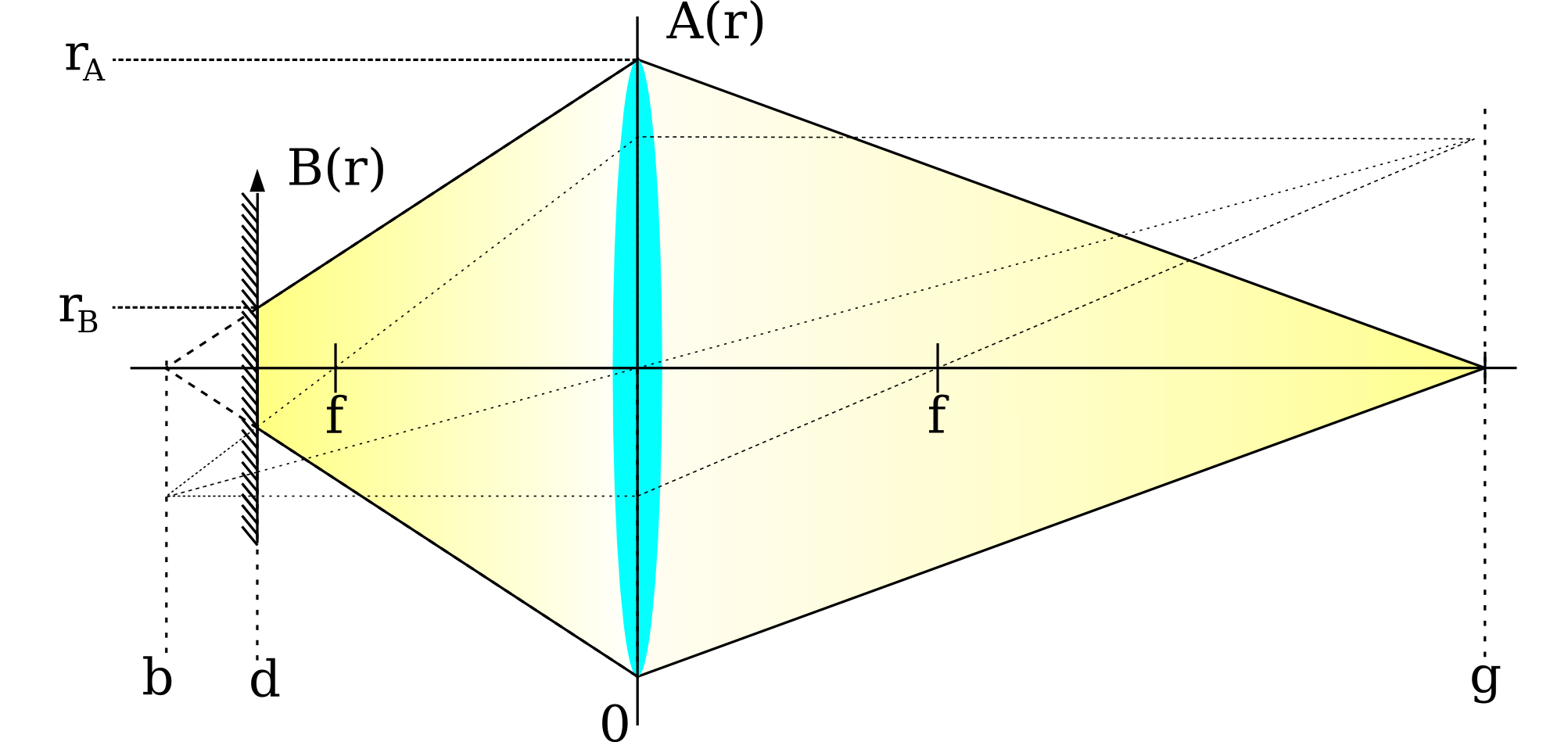}{%
A thin lens and an image sensor, facing an object where the image distance $\ImageDistance$ is not equal to the image sensor distance $\ImageSensorDistance$, $\ImageDistance \neq \ImageSensorDistance$. This is the \OutOfFocus{} situation.}{FigThinLensBokeh}{0.7} 
In this case, the point like object in distance $\ObjectDistance$ is not mapped sharply onto the image sensor plane in distance $\ImageSensorDistance$ and becomes blurred.
This intensity distribution of the object on the image sensor is called Bokeh function $\BokehFunction(r,\varphi)$.
In the case of thin imaging optics, this Bokeh $\BokehFunction(r,\varphi)$ is equal to the aperture function $\ApertureFunction(r,\varphi)$ up to a scaling factor in size $\frac{\BokehRadius}{\ApertureRadius}$
\begin{eqnarray}
    \label{EqBokehToAperture}
    \BokehFunction(r,\varphi) &=& \frac{\BokehRadius}{\ApertureRadius} \ApertureFunction(r,\varphi).
\end{eqnarray}
The Aperture function $\ApertureFunction(r,\varphi)$ is a two dimensional encoding of the transmission or reflection coefficient of the aperture.
Here $\varphi$ and $r$ are the polar coordinates orthogonal to the optical axis.
Given Figure \ref{FigThinLensBokeh} we use the intercept theorem and Equation \ref{EqThinLens} to find the scaling between $\BokehFunction$ and $\ApertureFunction$
\begin{eqnarray}
    \label{EqBokehRatio}
    \frac{\BokehRadius}{\ApertureRadius} &=& 1 - \ImageSensorDistance \left( \frac{1}{\FocalLength} - \frac{1}{\ObjectDistance} \right).
\end{eqnarray}
\subsection{How to observe the Bokeh of an imaging optic}
\label{SubSecBokehBringOut}
The image of an object plane in distance $\ObjectDistance$, which is mapped onto an \OutOfFocus{} image sensor in $\ImageSensorDistance$, is the result of a convolution of the object plane with the imaging system's Bokeh function $\BokehFunction(r,\varphi)$.
When there is only a single $\delta$-distribution like feature in the object plane, the image shows the convolution kernel itself, i.e. the Bokeh function $\BokehFunction(r,\varphi)$.
Therefore, to get an imaging optic's Bokeh $\BokehFunction$ one only needs to acquire images of a point like light source in an $\OutOfFocus$ configuration.
%
\subsection{The Bokeh of segmented imaging reflectors}
\label{SubSeqBokehOnSegmentedReflectors}
On segmented imaging reflectors the aperture function $\ApertureFunction$ has special features because of the shape of the mirror facets and the gaps in-between them.
Since the Bokeh function $\BokehFunction$ is a scaling of the aperture function $\ApertureFunction$, see Equation \ref{EqBokehRatio}, it will also have these special features.
However, this is only true when the individual facets are well aligned.\\
The Figures \ref{FigSegmentedReflector} and \ref{FigSegmentedReflectorBokeh} show both the same well aligned segmented imaging reflector.
The difference in the Figures is the image sensor position, which is an \InFocus{} set-up in Figure \ref{FigSegmentedReflector} and an \OutOfFocus{} set-up in Figure \ref{FigSegmentedReflectorBokeh}.
Figure \ref{FigSegmentedReflectorBokehMisaligned} shows that $\BokehFunction$ strongly depends on the alignment state of the reflector.
The light bundles of the individual facets might overlap, and thus lead to obvious mismatches between $\ApertureFunction$ and $\BokehFunction$.
\FigCapLab{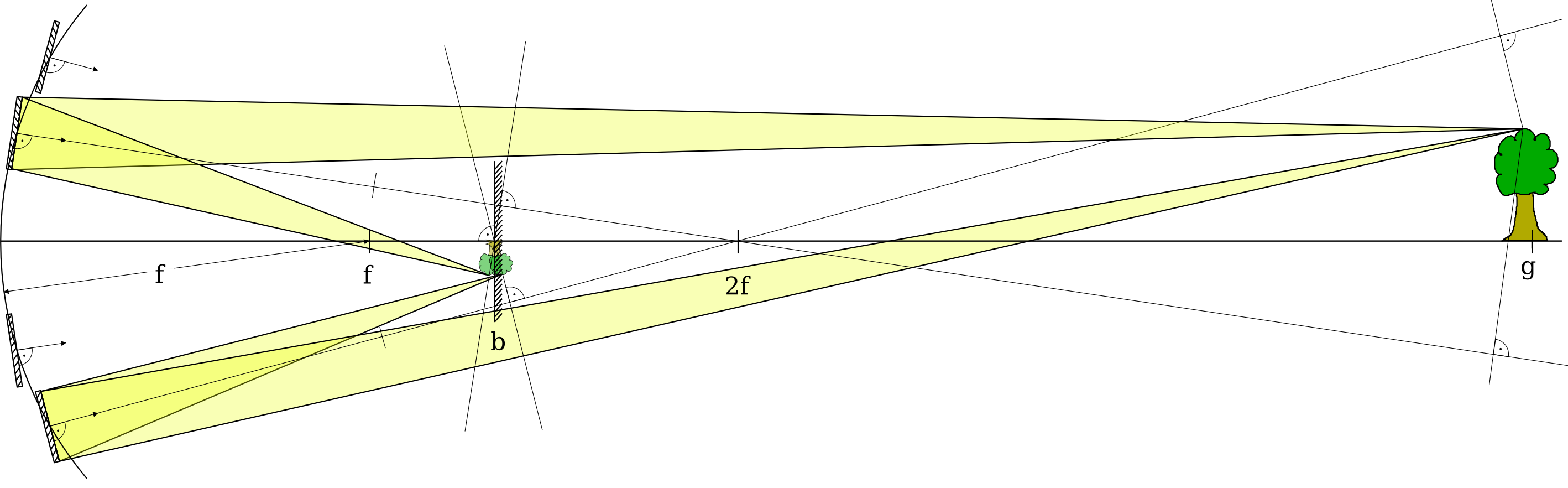}
{%
    A well aligned segmented imaging reflector in Davies Cotton \cite{Davies_Cotton_1957} configuration is observing a tree in distance $\ObjectDistance$. Only two of the four mirror light bundles are highlighted. The image sensor is placed in distance $\ImageSensorDistance=\ImageDistance$, which corresponds to the \InFocus{} situation of Equation \ref{EqThinLens}.
}{FigSegmentedReflector}
\FigCapLab{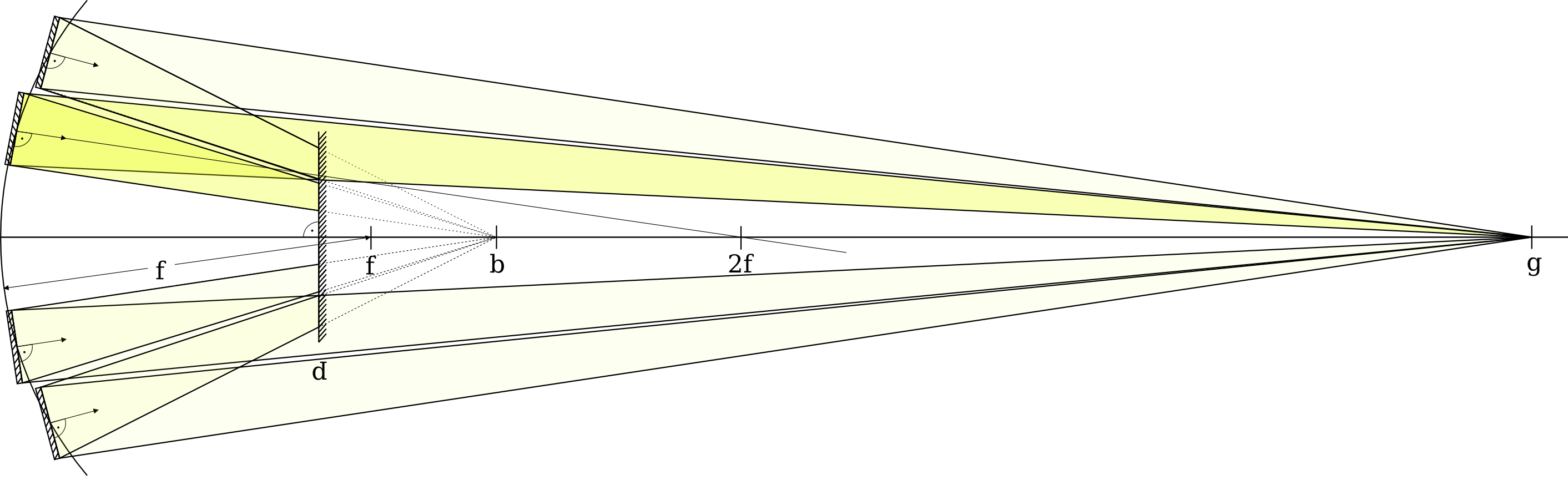}{%
    The segmented imaging reflector as in Figure \ref{FigSegmentedReflector}. A point like light source shines into the reflector while the image sensor is \OutOfFocus{} in $\ImageSensorDistance \neq \ImageDistance$. The image of the light source reveals the Bokeh $\BokehFunction$, see Section \ref{SubSecBokehBringOut}. All mirror light bundles are highlighted.
}{FigSegmentedReflectorBokeh}
\FigCapLab{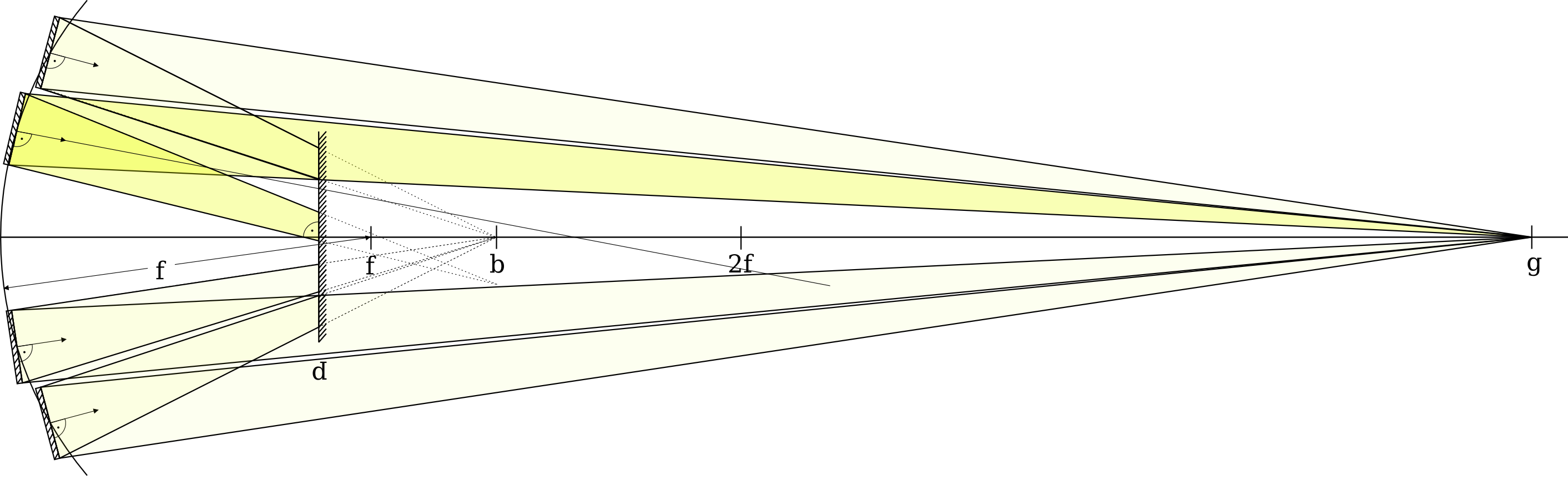}{%
    The segmented imaging reflector as in Figure \ref{FigSegmentedReflectorBokeh}, but now the second facet from the top is misaligned. $\BokehFunction$ does no longer match the scaled $\ApertureFunction$.
}{FigSegmentedReflectorBokehMisaligned}
%
\subsection{The $2\FocalLength$ method is a special case of Bokeh alignment}
\label{SubSec2fisSubSetOfBokeh}
In $2\FocalLength$ alignment, the light source or Bokeh lamp is placed in $\ObjectDistance = 2 \FocalLength$ and the image sensor to observe the mirror facet reflections is also placed in $\ImageSensorDistance = 2\FocalLength$.
According to Equation \ref{EqThinLens}, this is the $\InFocus$ configuration and according to Equation \ref{EqBokehRatio} the Bokeh ratio is $\frac{\BokehRadius}{\ApertureRadius} = 0$.
Therefore the observed Bokeh $\BokehFunction$ becomes a point like $\BokehRadius = 0$ spot observed during the $2\FocalLength$ alignment.
%
\section{How to perform Bokeh alignment on a segmented reflector}
\label{SeqHowTo}
We make use of the Bokeh's dependence on the alignment state of a segmented reflector to formulate an alignment procedure.
The facets can be aligned by matching the reflector's actual Bokeh $\BokehFunction$ to an ideal Bokeh template $\BokehTemplateFunction$, which is the Bokeh of the well aligned reflector.
In case of thin imaging systems, Equation \ref{EqBokehToAperture} shows us a simple way to obtain such a template $\BokehTemplateFunction$.
We just have to know the aperture $\ApertureFunction$ of the segmented reflector to calculate $\BokehTemplateFunction$ using the scaling Equation \ref{EqBokehRatio}. 
On non-thin imaging systems one can improve the $\BokehTemplateFunction$ by using different techniques, see Section \ref{sub_sec_ray_tracing_B_template}.
In general the following steps have to be done.
\begin{itemize}
    \item Create a Bokeh template $\BokehTemplateFunction$ of the well aligned segmented reflector.
    \item Place a screen on the reflector in distance $\ImageSensorDistance$ to project the reflector's Bokeh $\BokehFunction$ on. The screen should not cause additional shadowing, i.e. it should not be bigger than the image sensor housing of the telescope.
    \item Point the telescope to a point like light source in distance $\ObjectDistance$. We call this light source Bokeh lamp.
    \item Observe the projected Bokeh $\BokehFunction$ on the screen and compare it to the template $\BokehTemplateFunction$.
    \item Align each facet to make its own $\BokehFunction$ on the screen overlap with its $\BokehTemplateFunction$.
    \item When the reflector's Bokeh $\BokehFunction$ matches $\BokehTemplateFunction$, the reflector is aligned.
\end{itemize}
%
\section{The Bokeh alignment of \acs{fact}}
\label{SeqStepByStep}
We performed Bokeh alignment on \acs{fact} in May 2014 in the simple way explained in Section \ref{SeqHowTo}.
It should be noted that for the Bokeh alignment of \acs{fact}, not a single line of computer program was written. Further, the whole procedure, including preparation and hardware construction, took only a single day.
All the tools used were:
\begin{itemize}
    \item A Bokeh lamp, 3\,W LED
    \item A desktop printer
    \item A consumer image processing tool, GIMP on gnu-Linux
    \item A consumer photo camera, Nikon DSLR with telephoto lens
    \item A laser distance meter
    \item Water levels and laser cross line levels
\end{itemize}
Table \ref{TabFact} and Figure \ref{FigFACT} show the basic properties of \ac{fact}'s imaging system. 
\FigCapLab{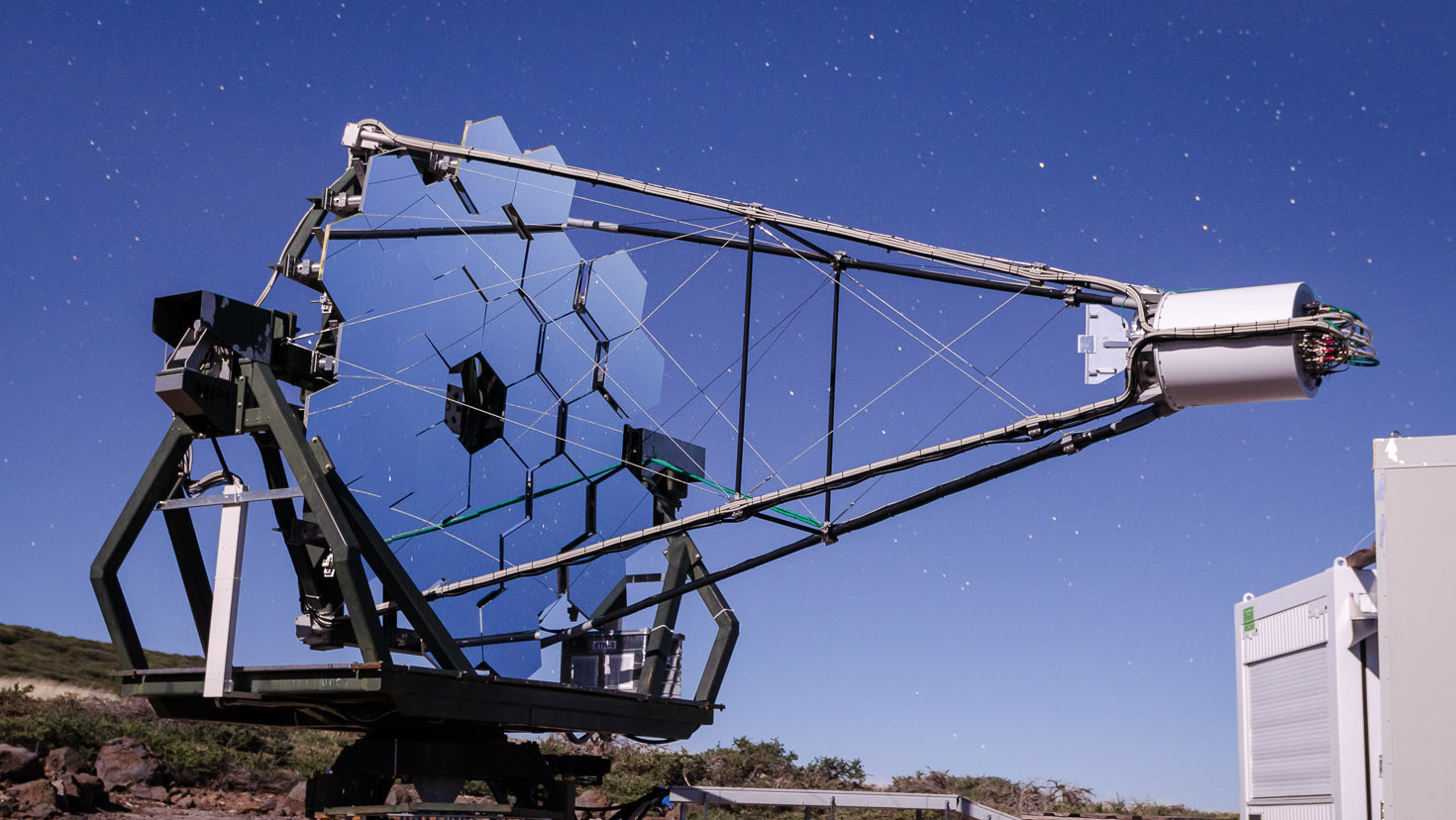}{%
    \acs{fact} is located on Canary island La Palma, Spain. It inherited its mount and the mirror facets from HEGRA \cite{HEGRA_status_and_results_1998}. While pioneering silicon photomultipliers for \acs{iact}s, \acs{fact} is monitoring $\gamma$ ray bright Blazars such as Mrk\,421 and Mrk\,501. Photograph by Thomas Kr\"ahenb\"uhl.
}{FigFACT}
\begin{table}[h]
    \begin{center}
        \begin{tabular}{lr}
            \toprule
            focal length $\FocalLength$ & $4.889\,$m\\
            number of facets & $30$\\
            facet mounting & manual adjustment on tripod\\
            reflector geometry along optical axis & $1/2$\,Davies Cotton + $1/2$\,parabola\\
            reflector area $A$ & $9.51\,$m$^2$\\
            effective reflector area $A_\text{eff}$ & $8.80\,$m$^2$\\
            effective aperture diameter $\ApertureDiameter_\text{eff}$ & $3.35\,$m\\
            maximum aperture diameter $\ApertureDiameter_\text{max}$ & $3.93\,$m\\
            effective F-number, $\FocalLength/\ApertureDiameter_\text{eff}$ & $1.46$\\
            F-number, $\FocalLength/\ApertureDiameter_\text{max}$ & $1.25$\\
            image sensor diameter, FoV & $0.39\,$m, $4.5^\circ\,$deg\\
            \bottomrule
        \end{tabular}
        \caption[]{Basic imaging reflector properties of \ac{fact}}
        \label{TabFact}
    \end{center}
\end{table}
%
\subsection{Creating the Bokeh template $\BokehTemplateFunction$ for the well aligned \acs{fact}}
First we need to know the Bokeh $\BokehTemplateFunction$ of the well aligned \ac{fact}.
We assume that \ac{fact}'s imaging system is thin, i.e. that the methods from Chapter \ref{SecBokehThin} are applicable.
It is important to note, that \ac{fact}'s reflector is neither thin in Davies Cotton nor in parabolic configuration.
In Section \ref{sub_sec_ray_tracing_B_template} we extend the method to non thin imaging systems.
For this first demonstration however, we restricted ourselves to the thin lens approximation, for the sake of simplicity.
Now we can, according to Equation \ref{EqBokehToAperture}, use the aperture function $\ApertureFunction$ to get the Bokeh template $\BokehTemplateFunction$.
In a distance $>10 \FocalLength$ of \acs{fact}, we take a picture of the reflector's aperture $\ApertureFunction$ with a telephoto lens for minimal distortions and measure $\ApertureFunction$ with maximum contrast, see the red part of Figure \ref{FigApertureImageHighlight}.
\FigCapLab{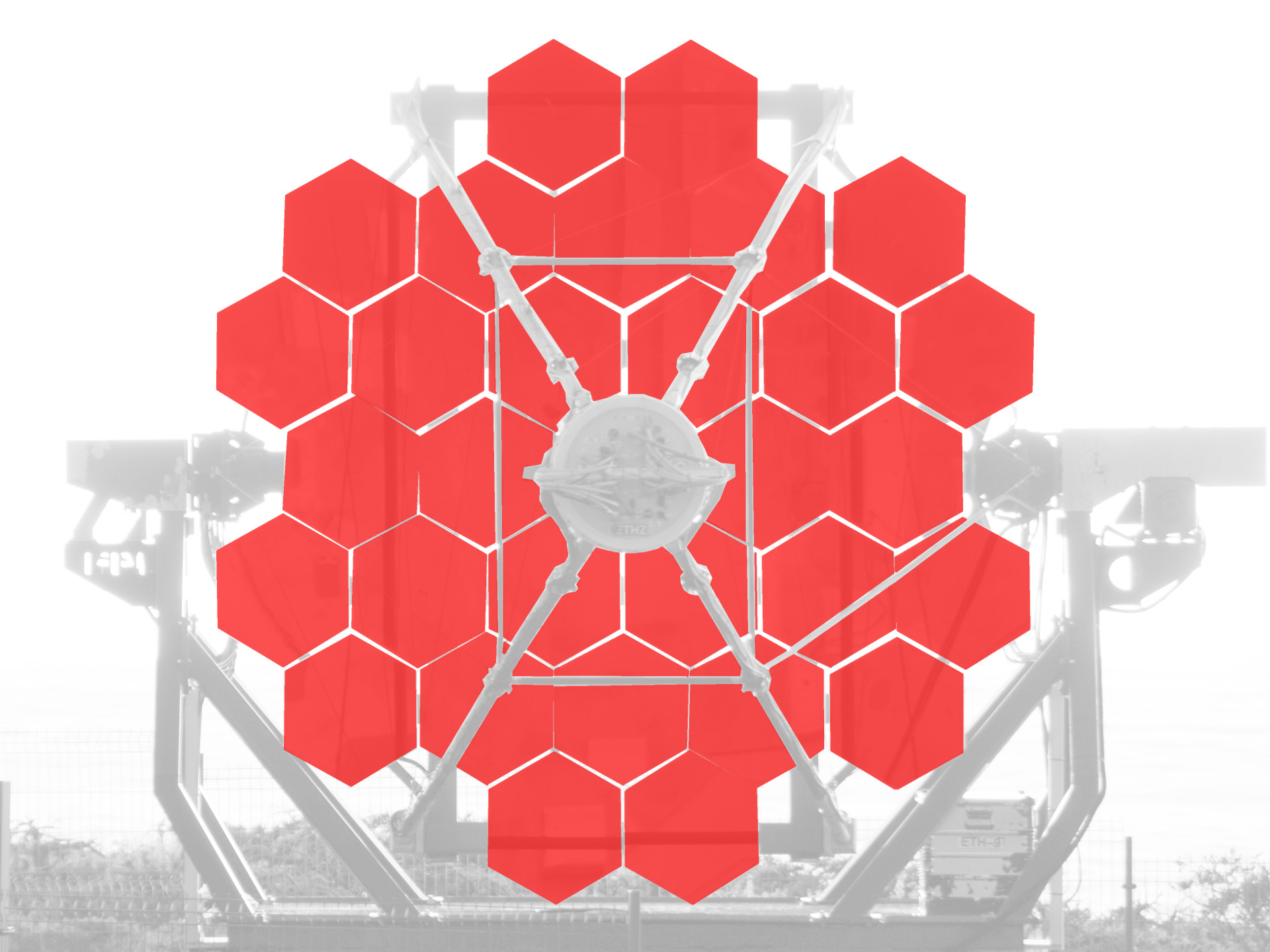}{%
    The aperture function $\ApertureFunction$ in red, extracted from the aperture photograph.
}{FigApertureImageHighlight}
%
\subsection{Creating the Bokeh template screen}
The Bokeh template $\BokehTemplateFunction$ is printed on paper and glued onto a flat shield.
This template screen serves as temporary imaging screen, see Figure \ref{FigBokehTemplateScreenMounting}.
The size of the screen $\BokehRadius$ is chosen such that the screen does not create additional shadowing on the reflector.
The screen is mounted on the telescope in distance $\ImageSensorDistance$ w.r.t. the reflector, which is $10\,$cm on top of \ac{fact}'s image sensor.
The Bokeh parameters of the alignment of \acs{fact} are shown in Table \ref{TabBokehParameters}.
\begin{table}[H]
    \center
    \begin{tabular}{lr}
        \toprule
        Aperture radius $\ApertureRadius$ & 1.633\,m\\
        Bokeh radius $\BokehRadius$ & 0.190\,m\\
        Bokeh scaling $\frac{\BokehRadius}{\ApertureRadius}$ & 0.1164\\
        $\BokehTemplateFunction$ distance $\ImageSensorDistance$ & 4.789\,m\\
        Bokeh lamp distance $\ObjectDistance$ & 49.939\,m\\
        \bottomrule
    \end{tabular}
    \caption[]{
        Bokeh parameters for the alignment of \acs{fact}. Given $\FocalLength$ and $\ApertureRadius$ from \acs{fact}'s design, we chose $\BokehRadius$ and $\ImageSensorDistance$ to fit our needs. Finally, we calculate $\ObjectDistance$ according to Equation \ref{EqBokehRatio}. }
    \label{TabBokehParameters}   
\end{table}
\FigCapLab{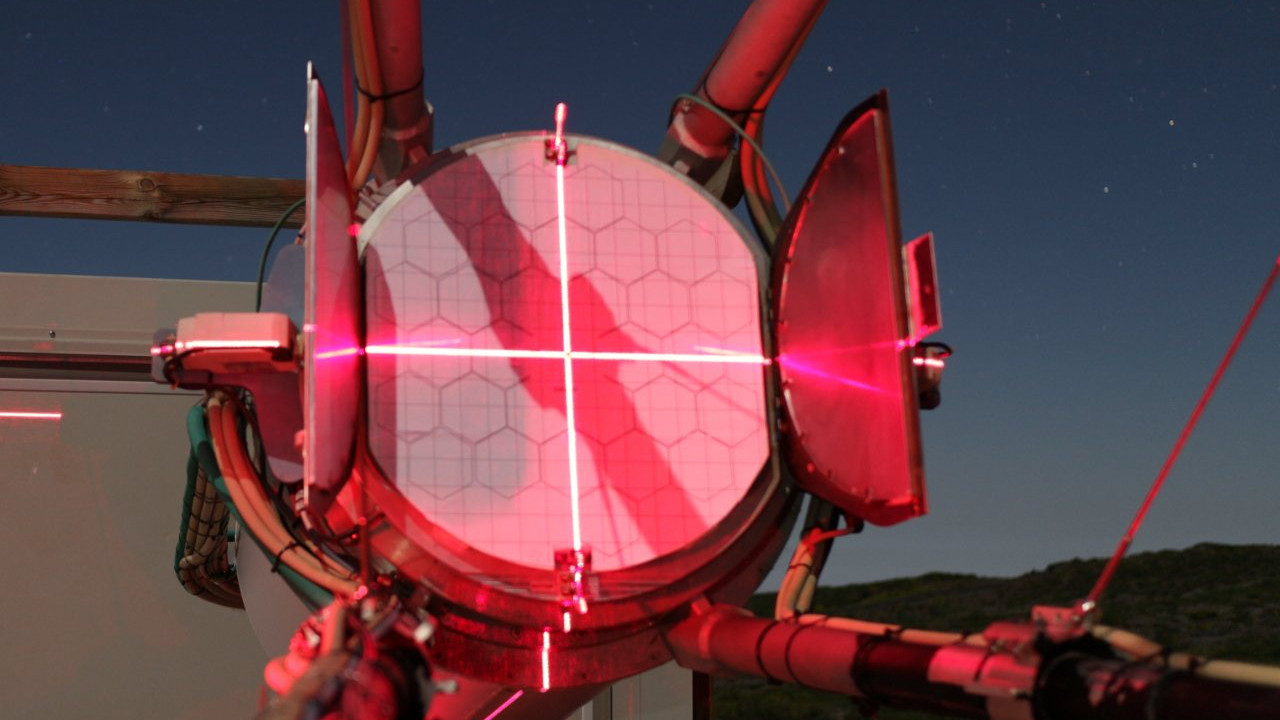}{%
    Mounting and positioning the $\BokehTemplateFunction$ screen in front of \acs{fact}'s image sensor using the laser cross line level.
}{FigBokehTemplateScreenMounting}
%
\subsection{Observing \acs{fact}'s Bokeh}
To observe the Bokeh $\BokehFunction$ from the currently misaligned reflector, we point the telescope to our Bokeh lamp, see Figure \ref{FigBokehAlignmentHalfDone}.
Our reflector then maps the actual $\BokehFunction$ onto the $\BokehTemplateFunction$ screen.
For crosscheck, the single mirror facet's $\BokehFunction$ now should have the same size as the individual facet templates in $\BokehTemplateFunction$.
%
\subsection{Aligning \acs{fact} using its Bokeh}
\label{SubSecAlignigFACT}
\begin{figure}[H]
    \begin{minipage}[t]{0.485\linewidth}
        \begin{figure}[H]
            \includegraphics[width=1.0\textwidth]{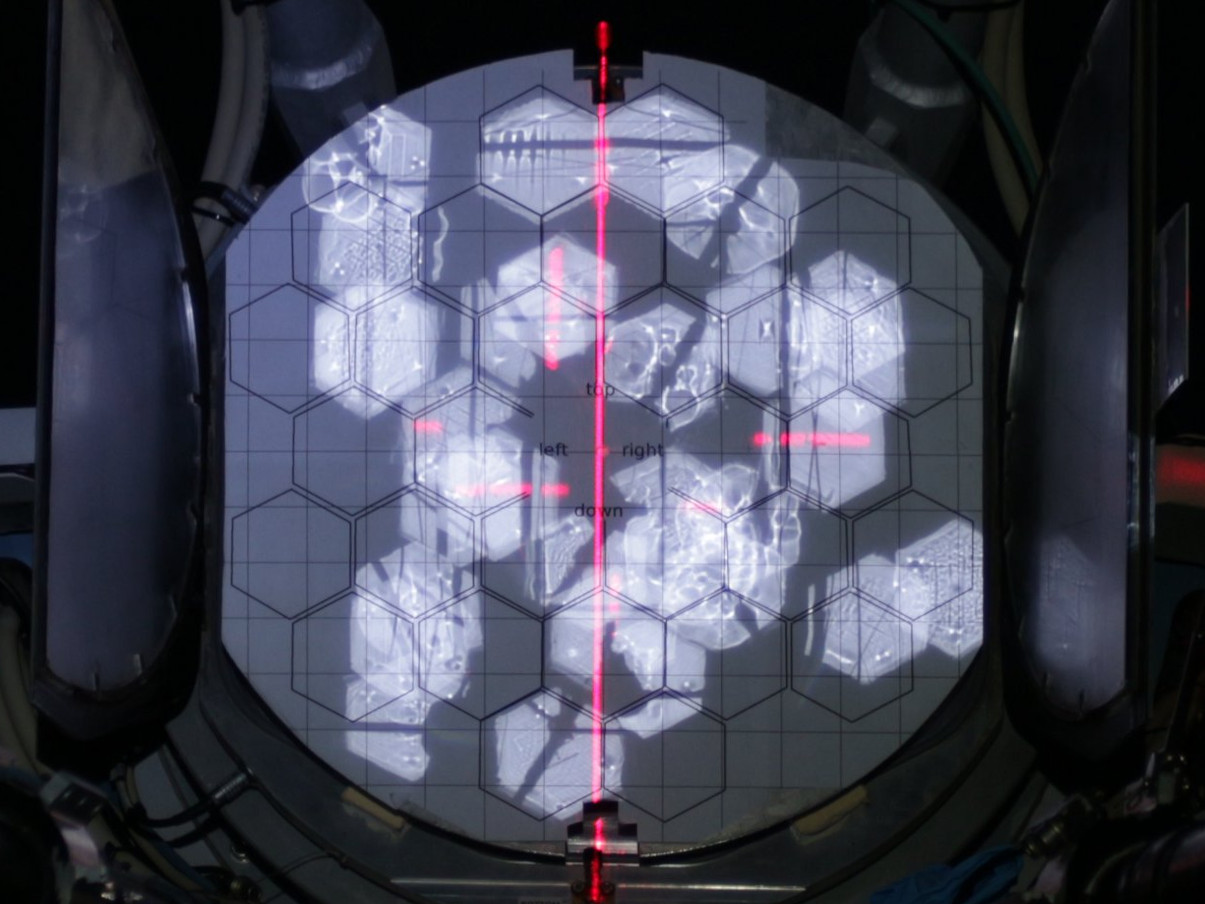}
            \caption[]{
                The initial Bokeh $\BokehFunction$ on \acs{fact} before alignment. It looks nothing like the template.
            }
            \label{FigBokehBeforeAlignment}
        \end{figure}
    \end{minipage}
    \hfill
    \begin{minipage}[t]{0.485\linewidth}
        \begin{figure}[H]
            \includegraphics[width=1.0\textwidth]{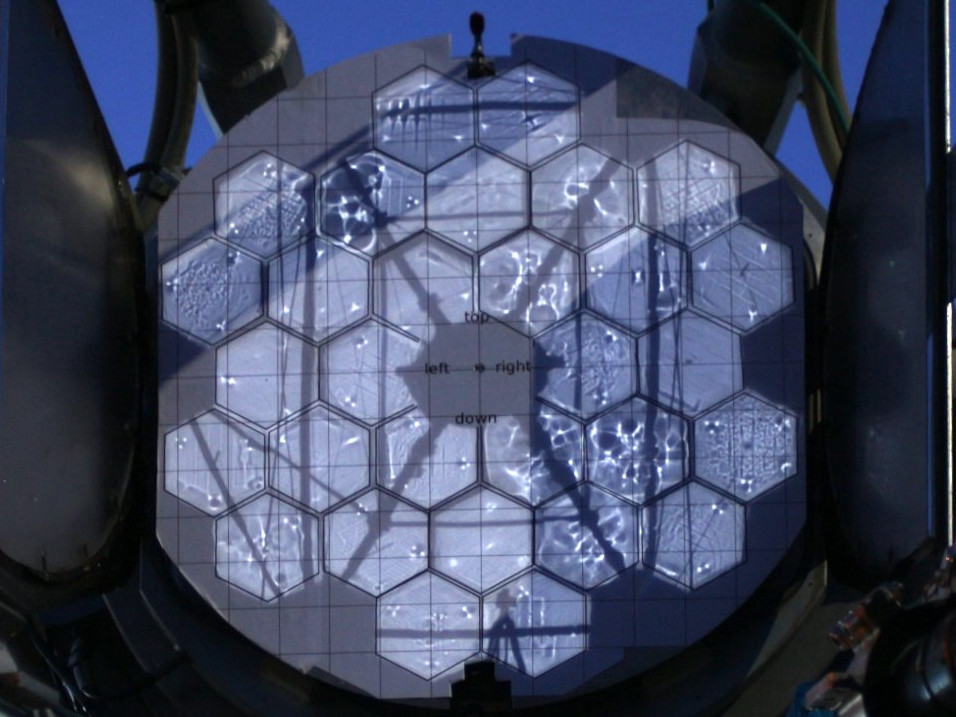}
            \caption[]{
                The Bokeh $\BokehFunction$  after finishing alignment. We finished alignment in late dawn. $\BokehFunction$ now matches the $\BokehTemplateFunction$, compare Figure \ref{FigApertureImageHighlight}.
            }
            \label{FigBokehAfterAlignment}
        \end{figure}
    \end{minipage}
\end{figure}

We reorient each mirror facet to make its individual $\BokehFunction$ overlap the corresponding one in $\BokehTemplateFunction$.
When the overlap is maximized, the alignment is done.
The mirror joints are operated manually while watching the $\BokehTemplateFunction$ screen and the $\BokehFunction$ on it.
Figures \ref{FigBokehAlignmentHalfDone}, \ref{FigBokehBeforeAlignment}, and \ref{FigBokehAfterAlignment} show the Bokeh alignment of \acs{fact} during, before, and after the alignment.
\FigCapLab{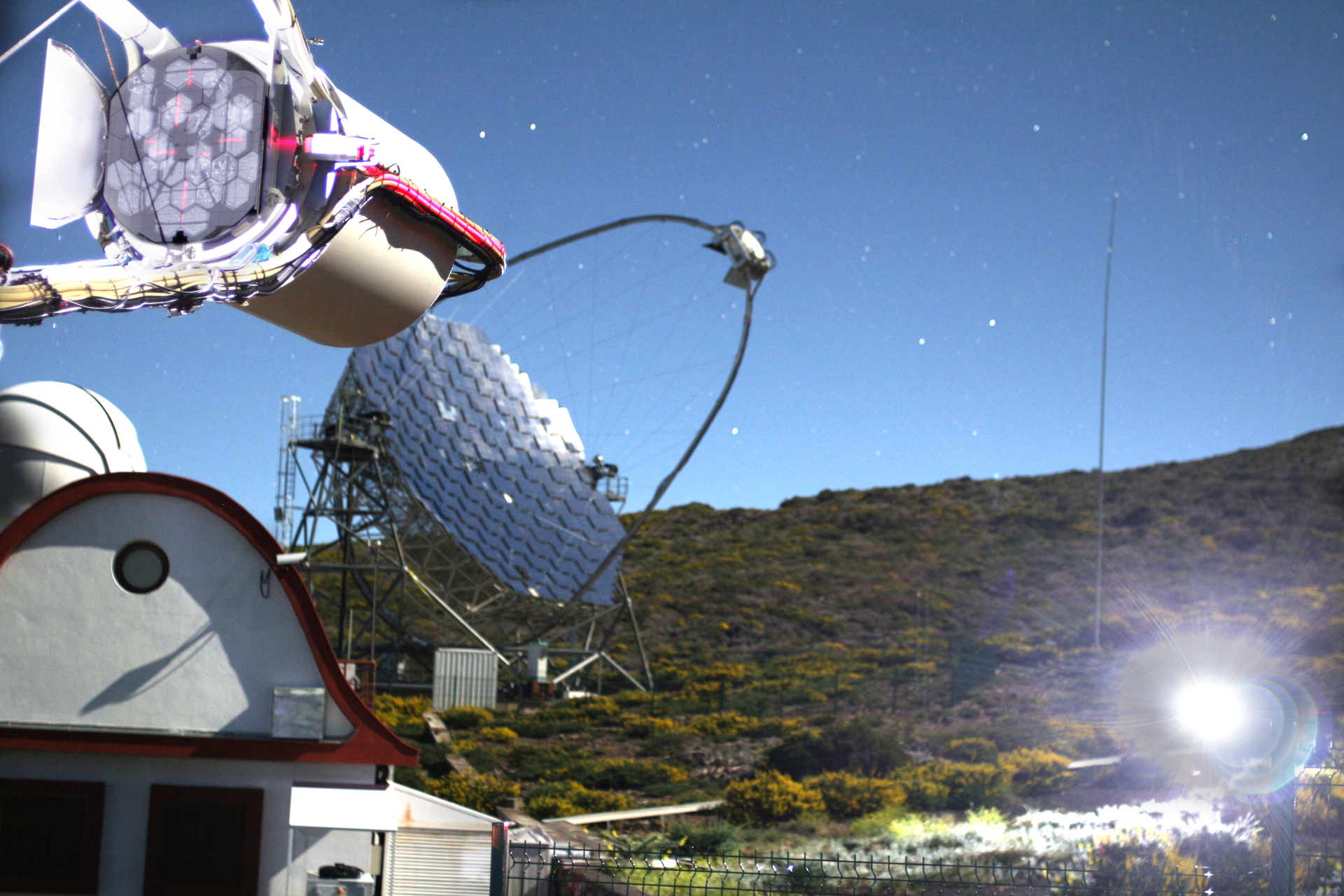}{%
    In the upper left we see \acs{fact}'s sensor housing and the Bokeh template screen on top of the imaging sensor. In the lower right part we see the Bokeh lamp. The lower half of $\BokehFunction$ already matches $\BokehTemplateFunction$. In the background, the MAGIC\,I telescope can be seen.
}{FigBokehAlignmentHalfDone}
%
\subsection{Investigating individual facets using Bokeh}
\label{SubSecIndividualFacets}
\begin{figure}
    \begin{minipage}[t]{0.485\linewidth}
        \begin{figure}[H]
            \includegraphics[width=1.0\textwidth]{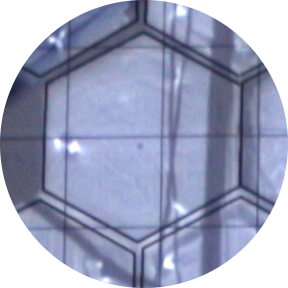}
            \caption[]{
                A homogeneous facet Bokeh. 4th row from top, 1st from left in figure \ref{FigBokehAfterAlignment}.
            }
            \label{FigFacetBokehGood}
        \end{figure}
    \end{minipage}
    \hfill
    \begin{minipage}[t]{0.485\linewidth}
        \begin{figure}[H]
            \includegraphics[width=1.0\textwidth]{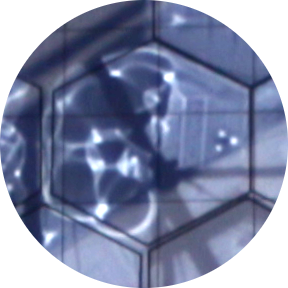}
            \caption[]{
               A partially distorted facet Bokeh. 2nd row from top, 2nd from left in Figure \ref{FigBokehAfterAlignment}.
            }
            \label{FigFacetBokehHeavyDistortion}
        \end{figure}
    \end{minipage}
\end{figure}

In the same way as the overall Bokeh of a segmented reflector tells about the alignment sate of its facets, the individual facet Bokeh tells about the properties of the facet's surface.
The Figures \ref{FigFacetBokehGood} and \ref{FigFacetBokehHeavyDistortion} show close up sections from Figure \ref{FigBokehAfterAlignment}.
Different types of surface distortions can be found. 
For example on \acs{fact}, the Bokeh reveals that all mirror facets have surface distortions at their tripod mounts.
In contrast to the more advanced Phase Measuring Deflectometry (PMD) technique \cite{CTA_PMD_ICRC2011}, a facet's Bokeh gives only limited information to reconstruct the facet's surface normals.
But the advantage of the Bokeh method is its simple execution so that facet deformations, e.g. due to aging, might be recognized on a daily basis while the facets are installed on the telescope.
Individual facet surface deformation information can not be deduced easily from a \acs{psf} image, where all individual facet \acsp{psf} are stacked on top of each other.
%
\subsection{Resulting Imaging performance on \acs{fact}}
\acs{fact} was initially aligned in 2011 using the $2f$ method \cite{FACT_design}.
The \acs{psf} achieved in the $2f$ method was in use until May 2014, and is shown in Figure \ref{FigPsfBeforeBokeh}.
This \acs{psf} lead in particular to the characterization of the $\gamma$ ray emission from the Crab nebula \cite{ICRC2014_fact_crab_spectrum}.
In May 2014 the reflector geometry was reconfigured to form a hybrid of Davies Cotton and parabola in order to improve the time resolution \cite{FACT_Muon_calibration_ICRC2015} of the reflector.
%
After this reconfiguration, \acs{fact} was unusable since every facet was arbitrarily oriented.
We performed the realignment using the novel Bokeh method.
The new \acs{psf} after Bokeh alignment is shown in Figure \ref{FigPsfAfterBokeh}.
Bokeh alignment achieved a \acs{psf} similar in size to the one before the reflector reconfiguration.
We did one Bokeh alignment iteration in one day.
The \acs{psf} images \ref{FigPsfBeforeBokeh} and \ref{FigPsfAfterBokeh} are recorded in the reflector's focal plane using a $6 \times 6\,$cm$^2$ radiometrically calibrated image sensor while \acs{fact} is tracking the star Arcturus.
This dedicated image sensor was made out of a vintage medium format camera (Hasselblad 6x6) and an industrial ccd camera.
Mounted on FACT, this image sensor covers $0.75^\circ \times 0.75^\circ$.
%
In both Figures \ref{FigPsfBeforeBokeh} and \ref{FigPsfAfterBokeh}, an enclosure ellipse of the light intensity distribution is highlighted in red together with a blue \acs{fact} pixel aperture for size comparison.
The \acs{psf} in Figure \ref{FigPsfAfterBokeh} is a bit blurred because of wind during the exposure.
The enclosure ellipse areas $A_{\sigma}$ are compared in Table \ref{TabSigmaArea}.
\begin{table}
    \begin{center}
        \begin{tabular}{lr}
            Reflector state &  $A_{\sigma}$ [arcmin$^2$]\\
            \toprule
            before reconfiguration & $62.0$\\
            after reconfiguration & too large to be recorded\\
            after Bokeh alignment & $65.5$\\
            \bottomrule
        \end{tabular}
        \caption[]{
            The \acs{fact} \acs{psf} 1$\sigma$ area before and after the reflector reconfiguration and Bokeh alignment, see Figures \ref{FigPsfBeforeBokeh} and \ref{FigPsfAfterBokeh}.
        }
        \label{TabSigmaArea}
    \end{center}
\end{table}
If the time for more complex alignment methods can be afforded, Bokeh alignment is still efficient to be performed in advance as it was done later for the \acs{namod} alignment of \acs{fact}. \cite{FACT_NAMOD_alignment}.  
\FigCapLab{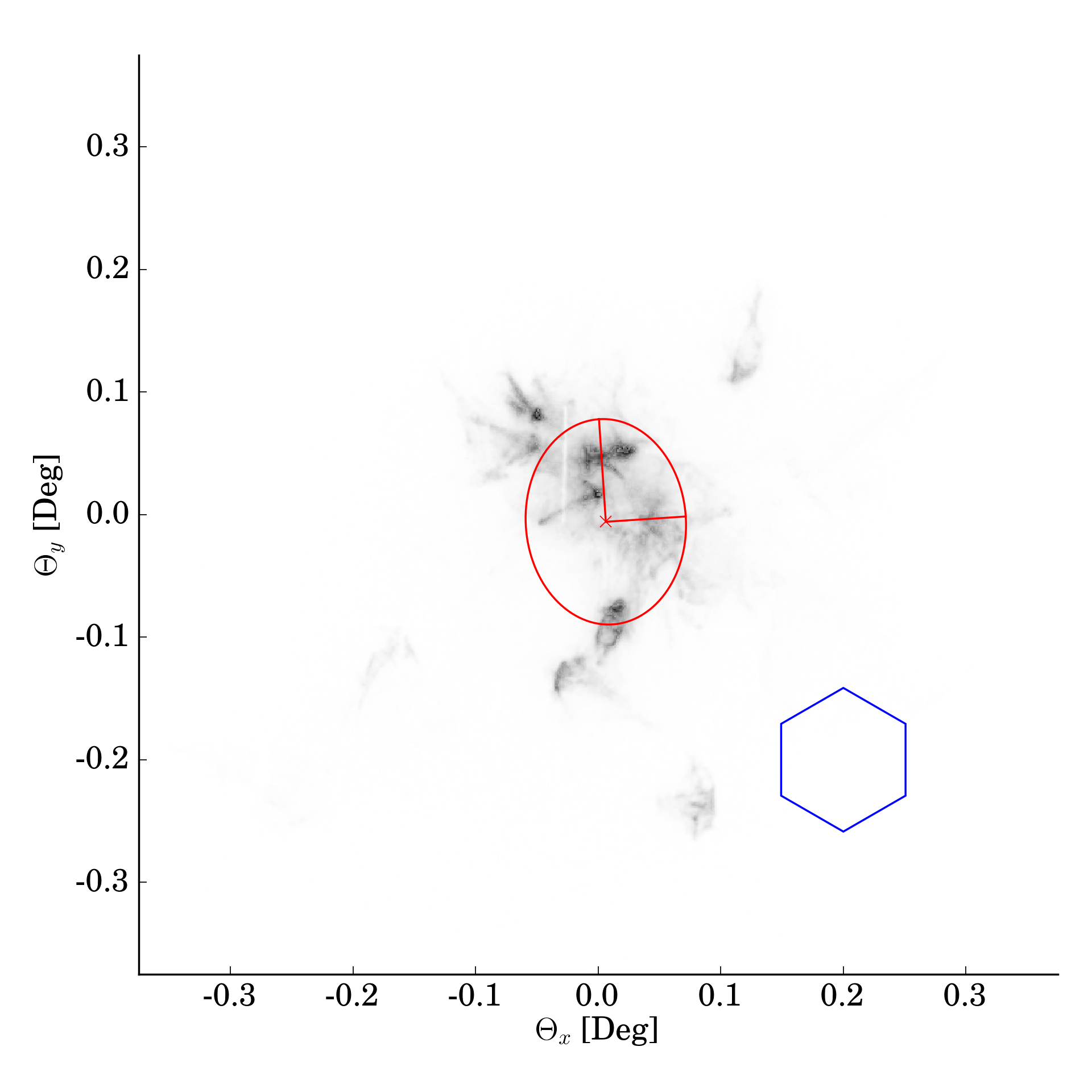}{%
    Image of star Arcturus in inverse color scale before the reflector reconfiguration. This \acs{psf} was achieved using the $2f$ method in 2011. The $1\sigma$ area \mbox{$A_{\sigma} =\,62.0\,$arcmin$^2$} is shown in red. For comparison, a \acs{fact} pixel is shown in blue.
}{FigPsfBeforeBokeh}
\FigCapLab{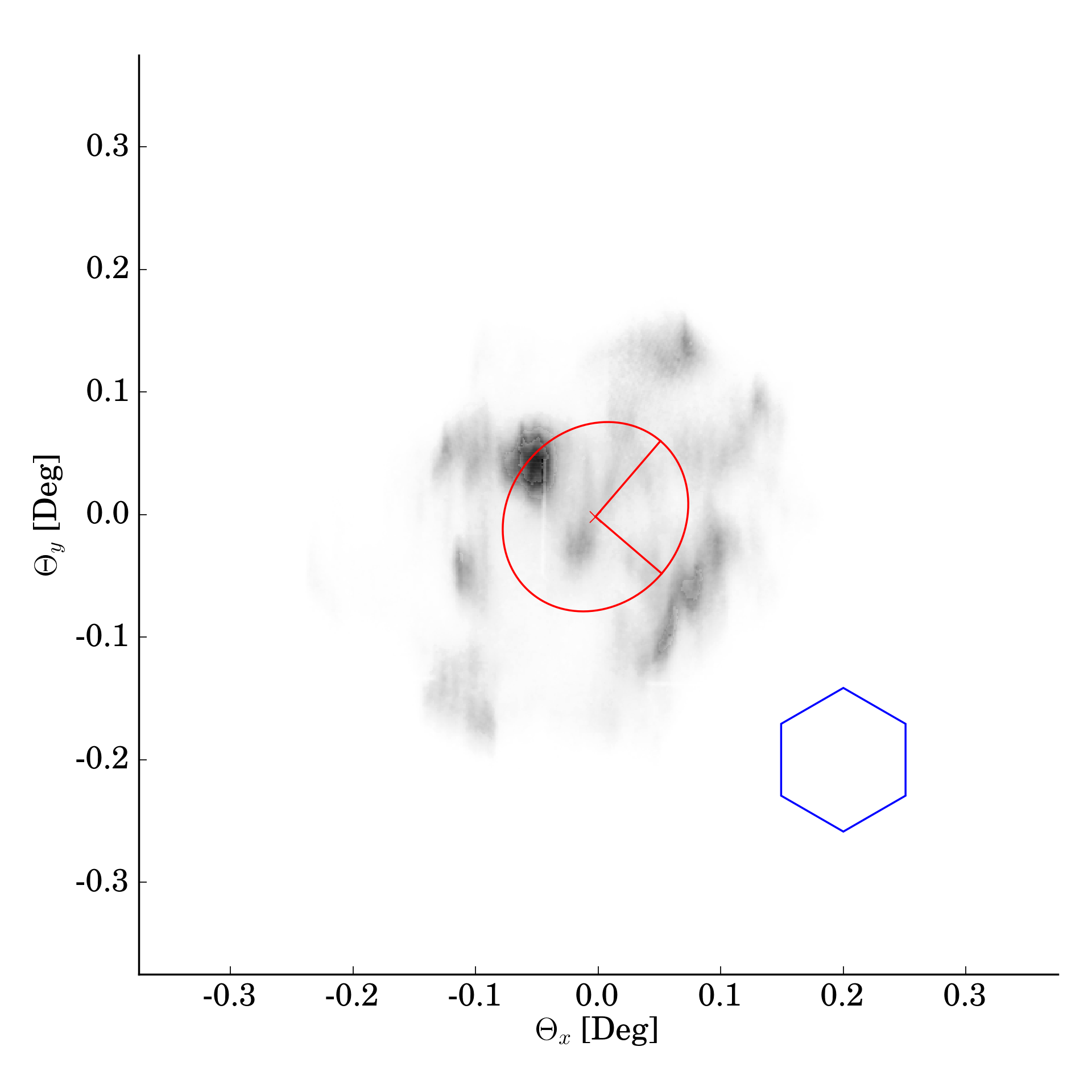}{%
    Image of star Arcturus in inverse color scale after reflector reconfiguration and Bokeh alignment. The $1\sigma$ area \mbox{$A_{\sigma} =\,65.5\,$arcmin$^2$} is shown in red. For comparison, a \acs{fact} pixel is shown in blue. Due to wind during the exposure, the \acs{psf} here does not show the same sharp features as they can be found in Figure \ref{FigPsfBeforeBokeh}.
}{FigPsfAfterBokeh}
%
\section{Outlook}
\label{SecOutlook}
In this outlook, we want to show what Bokeh alignment might look and feel like in the future, when one combines it with ray tracing simulations and computer vision \cite{Computer_Vision_Book_Forsyth_Ponce}.
%
\subsection{Dropping the thin lens approximation --- $\BokehTemplateFunction{}$ redone in ray tracing}
\label{sub_sec_ray_tracing_B_template}
Actual imaging systems are not thin and especially \ac{iact} optics are not thin with $\FocalLength/\ApertureDiameter$ below $2$.
After the first Bokeh alignment of FACT, we used ray tracing to obtain a better approximation of $\BokehTemplateFunction{}$ based on \acs{fact}'s actual non thin imaging reflector.
Figure \ref{FigSimulatedBokeh} shows the simulated light distribution on \acs{fact}'s Bokeh screen for the configuration stated in Table \ref{TabBokehParameters}.
The actually used $\BokehTemplateFunction{}$, created using the thin lens approximation and shown in Figure \ref{FigApertureImageHighlight}, is indeed slightly different from the ray tracing $\BokehTemplateFunction{}$.
For example the overlap of the facet edges is not predicted by the thin lens approximation.
Using this $\BokehTemplateFunction$ will result in a better \acs{psf} and is currently investigated on the \acf{cta} \cite{CTA_Introduction} Medium Size Telescope (MST) prototype in Berlin Adlershof.
\FigCapLabSca{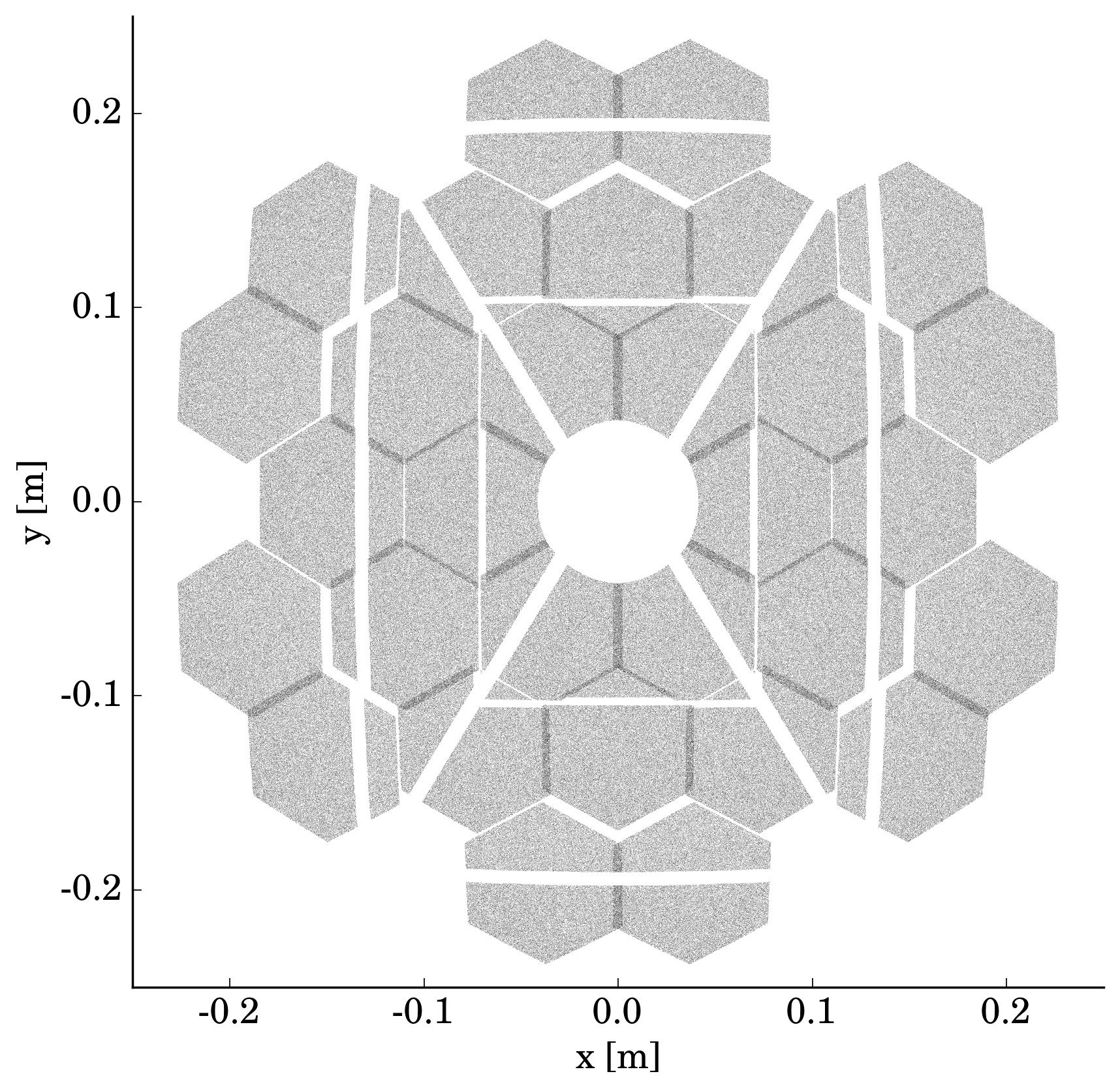}{%
    Simulated light distribution on the Bokeh screen of \acs{fact}, darker is more intense. Differences show up when comparing it to the thin lens approximated $\BokehTemplateFunction$ shown in Figure \ref{FigApertureImageHighlight}.
}{FigSimulatedBokeh}{0.85}
%
\subsection{A $\BokehTemplateFunction$ for off axis observation}
\label{SubSecOutlookOffAxis}
To increase further the flexibility, one could create $\BokehTemplateFunction$ for an off axis configuration using ray tracing simulation.
Here the Bokeh lamp is not restricted to the reflector's optical axis.
This way the inner facets of the reflector can be reached more easily in case the telescope's sensor housing is shadowing them. 
%
\subsection{Avoiding all facet ambiguity}
\label{SubSecOutlookAvoidingAmbiguity}
The problem of facet ambiguity is avoided when using a video projector instead of a simple lamp as Bokeh light source.
This way the projector can address individual facets without illuminating the rest, which eliminates all ambiguity in the facet Bokeh relation.
%
\subsection{Observing $\BokehFunction$ very fast}
\label{SubSecOutlookFastObservation}
\acs{fact}'s Bokeh alignment took several hours since we relied on pure eye hand coordination.
However, the observation of $\BokehFunction$ can also be done very quickly since all misalignment information of the facets can be deduced from a single photograph of $\BokehFunction{}$.
%
\subsection{Observing $\BokehFunction$ during the day}
\label{SubSecOutlookDuringDay}
The observation capabilities of $\BokehFunction$ depend only of the ratio of Bokeh lamp light intensity to ambient light intensity on the Bokeh screen.
For the night, dusk, and dawn, a 3\,W Bokeh lamp worked for \acs{fact} but during the day, a more powerful Bokeh lamp is needed.
We tested with a consumer flash light on \acs{fact} and have been able to record $\BokehFunction{}$ in direct sunlight.
%
\subsection{Alignment for all telescope elevations}
\label{SubSecGravitationalSlump}
On large telescopes gravitational slump effects the alignment depending on the elevation.
Since a Bokeh flash light is small it can be lifted to desired elevations rather easy using a floating balloon, fastened to the ground, or a multi copter drone.
Precise positioning of the Bokeh lamp is not crucial when the off axis Bokeh template is calculated and the lamp's position is reconstructed accurately.   
\subsection{Next level Bokeh alignment}
\label{SubSecOutlookNextLevel}
The lid of the telescope's image sensor will serve as Bokeh screen and will therefore be matt white.
A flying drone is given a remote controlled flash light as Bokeh lamp.
Additionally, drone cameras are mounted on the edge of the reflectors in order to reconstruct stereoscopically the position of the drone.
A lid camera, located in the reflector dish, is observing the lid screen.
A computer vision based Bokeh program reads out the lid camera, and the drone position.\\
During dusk, the drone is started.
Following the drone, the telescopes are moved to typical observation elevations.
The Bokeh flash light is fired by the Bokeh program and the resulting Bokeh on the lid screen is recorded by the lid camera.
Now the Bokeh program feeds the relative orientation of the drone and the telescope to a ray tracing simulation.
The ray tracing simulation then returns the $\BokehTemplateFunction$ for this particular configuration as proposed in Section \ref{sub_sec_ray_tracing_B_template} and \ref{SubSecOutlookOffAxis}.
From the deviations in between $\BokehTemplateFunction$ and $\BokehFunction$, the Bokeh program calculates the correction manipulations for the facets and sends these instructions to the mirror facet orientation control system of the telescope.\\
In such a setup one can:
\begin{itemize}
    \item align the telescope without star light, dark nights or good weather.
    \item align without any overlap with $\gamma$ ray observation time.
    \item reach an alignment accuracy that is not intrinsically limited by the thin lens approximation.
    \item acquire the facet orientations for any telescope elevation.
    \item have special $\BokehTemplateFunction$ for individual use cases.
    \item keep track of each facet surface conditions.
    \item align a whole array of \acs{iact}s with a single flying drone.
\end{itemize}
\section{Conclusion}
In a first demonstration we aligned the \acs{fact} \acs{iact} using Bokeh alignment in a single day and a single iteration from an useless state, after a reflector reconfiguration, to an operational state for $\gamma$ ray astronomy.
Bokeh alignment is inexpensive, fast and more versatile than current methods for the alignment of a segmented reflector.
Boke alignment is simple, and does not need a lot of custom hardware or software.
Yet it gives information about individual mirror facet surfaces and does neither need star light nor dark nights to align a segmented reflector.
Boke alignment could even be done during the day.\\
Therefore, we believe that Bokeh alignment is ideal to align a large number of Cherenkov telescopes, for instance the upcoming \acf{cta}.
%
\section*{Acknowledgments}
The  important  contributions  from  ETH Zurich  grants  ETH-10.08-2  and  ETH-27.12-1  as  well  as  the funding by the German BMBF (Verbundforschung Astro- und Astroteilchenphysik) are gratefully acknowledged.
We are thankful for the very valuable contributions from E.~Lorenz, D.~Renker and G.~Viertel during the early phase of the project.
We thank the Instituto de Astrofisica de Canarias allowing us to operate the telescope at the Observatorio Roque de los Muchachos in La Palma, the Max-Planck-Institut fuer Physik for providing us with the mount of the former HEGRA CT 3 telescope, and the MAGIC collaboration for their support.
We would also like to thank Karsten Schuhmann for discussions on how to address individual mirror facets with a projector and fruitful discussions on optics in general.
\section*{References}
\bibliography{references.bib}
\begin{acronym}
    \acro{sccan}[SCCAN]{Solar Concentrator Characterization At Night}
    \acro{psf}[PSF]{Point Spread Function}
    \acro{fact}[FACT]{First Geiger-mode Avalanche Cherenkov Telescope}
    \acro{veritas}[VERITAS]{Very Energetic Radiation Imaging Telescope Array System}
    \acro{iact}[IACT]{Imaging Atmospheric Cherenkov Telescope}
    \acro{namod}[NAMOD]{Normalized and Asynchronous Mirror Orientation Determination}
    \acro{cta}[CTA]{Cherenkov Telescope Array}
\end{acronym}
\end{document}